\documentclass[twocolumn,aps,prb]{revtex4-1}
\usepackage{graphicx}
\usepackage{times}
\usepackage[novolumeabbr]{unitsdef}
\usepackage{subfigure}
\usepackage{ulem}
\usepackage{color}
\usepackage{amsmath}

\begin{document}
\title{Interfacial effects revealed by ultrafast relaxation dynamics in BiFeO$_{3}$/YBa$_{2}$Cu$_{3}$O$_{7}$ bilayers}
\author{D. Springer}
\author{Saritha K. Nair}
\author{Mi He}
\author{C. L. Lu}
\altaffiliation{Present address: School of Physics, Huazhong University of Science and Technology, Wuhan 430074, China}
\author{S. A. Cheong}
\author{T. Wu}
\altaffiliation{Present address: Physical Sciences and Engineering Division, King Abdullah University of Science and Technology, Thuwal, 23955-6900, Saudi Arabia}
\author{C. Panagopoulos}
\affiliation{Division of Physics and Applied Physics, School of Physical and Mathematical Sciences, Nanyang Technological University, Singapore 637371, Singapore}
\author{Jian-Xin Zhu}
\email{jxzhu@lanl.gov}
\affiliation{Theoretical Division and Center for Integrated Nanotechnologies, Los Alamos National Laboratory, Los Alamos, New Mexico 87545, USA}
\author{Elbert E. M. Chia}
\email{elbertchia@ntu.edu.sg}
\affiliation{Division of Physics and Applied Physics, School of Physical and Mathematical Sciences, Nanyang Technological University, Singapore 637371, Singapore}

\date{\today}

\begin{abstract}
The temperature dependence of the relaxation dynamics in the bilayer thin film heterostructure composed of multiferroic BiFeO$_{3}$ (BFO) and superconducting YBa$_{2}$Cu$_{3}$O$_{7}$ (YBCO) grown on (001) SrTiO$_{3}$ substrate is studied by time-resolved pump-probe technique, and compared with that of pure YBCO thin film grown under the same growth conditions. The superconductivity of YBCO is found to be retained in the heterostructure. We observe a speeding up of the YBCO recombination dynamics in the superconducting state of the heterostructure, and attribute it to the presence of weak ferromagnetism at the BFO/YBCO interface as observed in magnetization data. An extension of the Rothwarf-Taylor model is used to fit the ultrafast dynamics of BFO/YBCO, that models an increased quasiparticle occupation of the ferromagnetic interfacial layer in the superconducting state of YBCO.

\end{abstract}

\maketitle
\section{Introduction}
Perovskite oxides exhibit many emergent physical properties such as high-temperature superconductivity, colossal magnetoresistance, ferroelectricity, and magnetoelectric effects. Since many of these oxides are epitaxially compatible, it is possible to create heterostructures at the atomic level with unique properties.~\cite{Goldman2006} The capacity to fabricate multilayered structures whose component parts host interacting electrons, allows experimentalists to encourage strong correlations in combination with access to new symmetries and electronic band structures provided by the interfaces.~\cite{Panagopoulos2010,Panagopoulos2012}

Complex oxide heterostructures of superconductors have been studied extensively in recent years.~\cite{Ramesh1991,Pepe2009,Goldman1999} Bringing different transition metal cations with their localized $d$-electron physics and interacting charge, spin, and lattice degrees of freedom, into contact in a tunable crystalline environment, activates new electronic properties not observable in bulk compounds. Epitaxial strain mismatch, atomic coordination frustration, ordered spin and orbital states, charge flow across the interface, chemical frustration, and competing phases placed in close proximity conspire for new phases to emerge. These include novel magnetic, superconducting, and multiferroic phases potentially capable of providing innovative science and technology platforms. However, most effort has focused on quasi-equilibrium properties. To explore the emergence of unique transient states one should also investigate the dynamical degrees of freedom. 

Time-resolved pump-probe ultrafast optical spectroscopy has been widely used in probing the relaxation dynamics of photoexcited quasiparticles in correlated electron systems, and in particular, high-temperature superconductors (HTSCs). It has proven to be a useful time-domain tool to probe dynamics of complex structures~\cite{Pepe2009,Taneda2007,Liu2005,You2005,Pan2008} and coexisting~\cite{Nair2010}/competing~\cite{Chia2007,Chia2010} phases in superconductors. Thus ultrafast optical spectroscopy is well suited to study competing interactions in heterostructures containing HTSCs. Various pump-probe studies have been done on nanocomposites involving superconductors. For example, in superconductor-ferromagnet (S/F) bilayers such as Nb/NiCu, NbN/NiCu and YBCO/Au/NiCu, the initial fast relaxation, which involves Cooper-pair breaking and quasiparticle (QP) recombination, is always faster in S/F bilayers than in pure superconductors, but only in the superconducting state.~\cite{Pepe2009} The faster relaxation was attributed to the presence of suppressed superconductivity at the pump spot, called the hotspot, whose size is of the order of the superconducting coherence length $\xi$. These hotspots, where the superconducting energy gap $\Delta$ is reduced, act as energy traps that cause faster QP recombination and hence speed up the relaxation process.~\cite{Parlato2013} In cuprate-manganite S/F bilayers such as YBa$_{2}$Cu$_{3}$O$_{7}$/La$_{1-x}$Sr$_{x}$MnO$_{3}$ (YBCO/LSMO), the LSMO layer, due to the proximity effect, acts as an effective sink layer for QPs coming from YBCO, thus also resulting in a faster relaxation. Various x-ray and neutron techniques have been used to study bilayer/trilayer/superlattice HTSC-manganite heterostructures, such as YBCO/LCMO (La$_{1-x}$Ca$_{x}$MnO$_{3}$) and YBCO/LSMO.~\cite{Chakhalian2006,Zhang2009,Hoffmann2005,Uribe-Laverde2013,Satapathy2012} These studies reveal (a) a strongly (but not totally) suppressed Mn ferromagnetic moment at the manganite side of the interface, and/or (b) a small ferromagnetic Cu moment developing on the HTSC side of the interface. This is consistent with the observation of a reduced saturation magnetization
per manganite layer compared to the bulk value.~\cite{Zhang2009,Hoffmann2005} The suppression of ferromagnetic order at the interface was attributed to charge transfer across the interface that increases the electron occupancy at the Mn sites, and change of the manganite ground state from ferromagnetic to antiferromagnetic.~\cite{Stadler2000,Hoffmann2005}

Going beyond HTSC/manganite, it would be interesting to study HTSC/multiferroic nanocomposites such as YBCO/BFO (BiFeO$_{3}$). BFO is one of the most studied multiferroic materials because of its room temperature ferroelectricity (Curie temperature, $T_{C}$$\sim$1103~K)~\cite{Smolenskii1963} and antiferromagnetism (N\'{e}el temperature, $T_{N}$$\sim$643~K).~\cite{Fischer1980} Weak ferromagnetism has been observed in BFO thin films.~\cite{Wang2003} Multiferroics and HTSCs are by themselves complex materials with different and often antagonistic order parameters (magnetic \& superconducting). Magnetization data have shown weak ferromagnetism to coexist with superconductivity in BFO/YBCO bilayers.~\cite{Werner-Malentoa2009} An interesting question would be: in the bilayer heterostructure containing multiferroic BFO and HTSC YBCO, how does the interplay of various long-range orders in the constituent materials affect its dynamics? In particular, does the multiferrocity and/or the weak ferromagnetism in BFO, affect the QP relaxation in YBCO? Would the weak ferromagnetism of BFO play the same role as the ferromagnetism of LCMO in YBCO/LCMO bilayers, speeding up the recombination dynamics of YBCO in the superconducting state? In this Article, we report time-resolved pump-probe data on a multiferroic/superconductor BFO/YBCO bilayer thin film. We also observe a speeding up to YBCO dynamics in the BFO/YBCO bilayers as in YBCO/LSMO bilayers. Magnetization data reveal the presence of weak ferromagnetism in our bilayer. The speeding up of YBCO dynamics in the BFO/YBCO bilayers could hence be related to BFO overlayer acting as a trapping layer and/or to the interaction of weak ferromagnetism in BFO and superconductivity in YBCO. We used an extension of the Rothwarf-Taylor model to fit our pump-probe data.

The paper is organized as follows. In Sec. II, we present the experimental details on sample synthesis and characterization, and on the pump-probe measurement.
The latter is the focus of the present paper. In Sec. III A, we discuss the pump-probe data of YBCO thin film, and analyze it using the Rothwarf-Taylor model.
In Sec. III B, we discuss the pump-probe data of BFO-YBCO bilayer thin film, and analyze it using an extended Rothwarf-Taylor model. A summary is given in Sec.
IV.

\section{Experiment}
The bilayer thin-film heterostructure of BFO and YBCO was deposited on (001) oriented single-crystal SrTiO$_{3}$ (STO) (CrysTec GmbH, Berlin Germany) substrate
by pulsed laser deposition (PLD) technique using a 248~nm KrF excimer laser. The YBCO bottom layer with a thickness of $\sim$100~nm was grown in 150~mTorr
oxygen at a substrate temperature of 780$^{\circ}$C, and then the BFO thin film ($\sim$130~nm thickness) was deposited on top of the YBCO layer at 80~mTorr and
670$^{\circ}$C. YBCO thin films were deposited first since the high growth temperature ($\sim$780$^{\circ}$C) may degrade the BFO thin films. Here, two YBCO thin
films are grown at the same conditions. Subsequently, BFO is deposited on one of the films while the other YBCO film is used for comparison of the dynamics.

\begin{figure} \centering
\includegraphics[width=8cm]{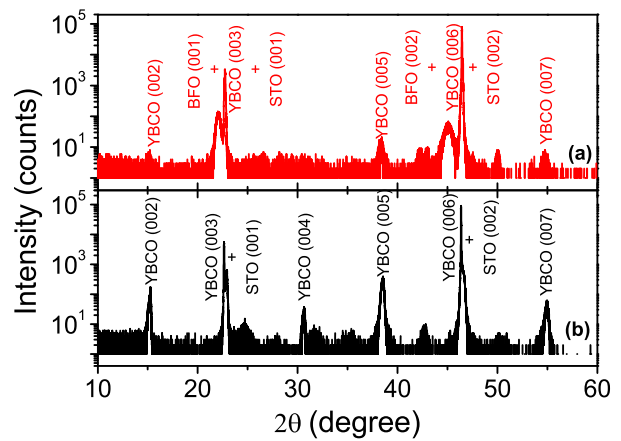}
\caption{X-ray diffraction patterns of (a) BFO/YBCO bilayer and corresponding (b) YBCO thin film.}
\label{fig:XRD}
\end{figure}

\begin{figure}
\centering
\includegraphics[width=8.5cm]{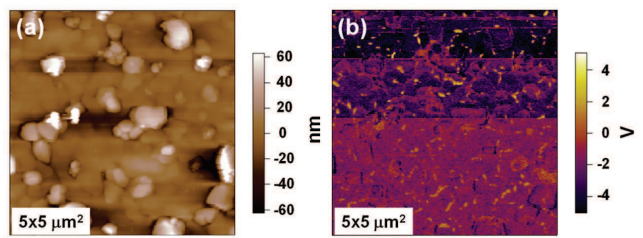}
\caption{(a) 5$\times$5~$\mu$m$^{2}$ topographic AFM image of the BFO (130~nm) grown on
YBCO/STO, and (b) corresponding piezoresponse phase image. Yellow (bright) and purple (dark)
correspond to the original up and down domains, respectively.}
\label{fig:AFM}
\end{figure}

\begin{figure}[ht]
\centering
\includegraphics[width=7cm]{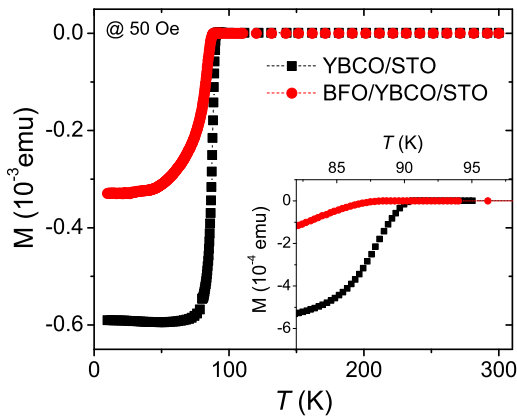}
\caption{Magnetization vs temperature measured on BFO/YBCO bilayer and corresponding YBCO thin film. }
\label{fig:SQUID}
\end{figure}

\begin{figure} \centering
\includegraphics[width=8.5cm]{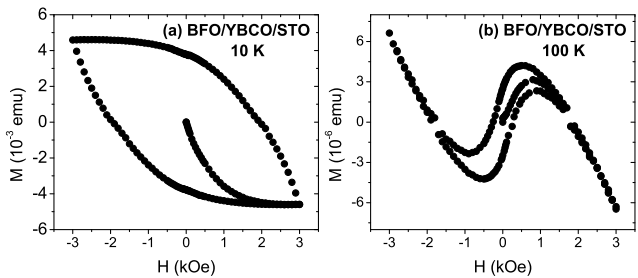}
\caption{Magnetization hysteresis loops of BFO/YBCO bilayer recorded (a) below and (b) above, $T_{c}$.}
\label{fig:MH}
\end{figure}

The structure and crystalline quality of synthesized thin films were studied by a high resolution x-ray diffractometer (Smartlab, Rigaku Corporation, Japan).
Figure~\ref{fig:XRD} displays the x-ray diffraction spectrum for the BFO/YBCO bilayer and corresponding YBCO thin film. No impurity phase was observed but only
(00l)-type diffraction peaks indicating the epitaxial growth of samples. The topographic imaging and the piezoresponse force microscopy (PFM) measurements were
carried out on an atomic force microscope (AFM) (Asylum Research, USA, MFP-3D). The 5$\times$5~${\mu}$m$^{2}$ topographic AFM image and the corresponding
out-of-plane PFM image are depicted in Figs.~\ref{fig:AFM}(a) and~\ref{fig:AFM}(b), respectively. Due to the strain relaxation, the surface morphology of YBCO thin
films is not smooth enough leading to a root mean square roughness (R$_{rms}$) of $\sim$19~nm of BFO/YBCO bilayer thin film. In a PFM image, yellow (bright)
and purple (dark) contrasts  represent spontaneous polarization component pointing `up' and `down', respectively.~\cite{Lu2010} Hence Fig.~\ref{fig:AFM}(b),
with the purple domain dominating, indicates that majority of the ferroelectric domains are pointing `down' for the BFO/YBCO bilayer sample, which is in
agreement with reported `down' polarization in BFO/YBCO bilayers.~\cite{Crassous2011} Further, we measured the $T_{c}$ of our samples by taking magnetization
data using Magnetic Property Measurement System (MPMS, Quantum Design). The temperature-dependent magnetization of BFO/YBCO bilayer and YBCO films are depicted
in Fig.~\ref{fig:SQUID}, where a magnified view near the transition demonstrates the onset of superconductivity at $T_{c}$$\sim$90~K in both films. It shows
that after depositing the BFO thin film on YBCO/STO, $T_{c}$ does not change. Figure~\ref{fig:MH} shows the magnetization hysteresis loops of BFO/YBCO bilayer
recorded below and above $T_{c}$ in magnetic field parallel to the sample surface. The data shows superconducting response at low temperature and a weak
ferromagnetic response above $T_{c}$, which also agrees with our recent predictions based on first-principles simulations.~\cite{Zhu:2014aa} No weak ferromagnetic response was observed in the YBCO thin film.

In our pump-probe experimental setup in reflectance geometry, an 80~MHz Ti:Sapphire laser produces sub-50 fs pulses at $\approx$800~nm (1.55~eV) as a source of
both pump and probe pulses. The pump and probe pulses were cross polarized. The pump spot diameter was 60~${\mu}$m and that of the probe was 30~${\mu}$m. The
reflected probe beam was focused onto an avalanche photodiode detector. The photoinduced change in reflectivity (${\Delta}R/R$) was measured using lock-in
detection. In order to minimize noise, the pump beam was modulated at 100~kHz with an acousto-optical modulator. The experiments were performed with an average
pump power of 2~mW, giving a pump fluence of $\sim$0.9~${\mu}$J/cm$^{2}$. The probe intensity was approximately ten times lower. Data were taken from 10~K to
120~K. In all the data, the temperature increase of the illuminated spot has been accounted for. The optical penetration depth of 800~nm laser in YBCO is
$\sim$100~nm,~\cite{Li2004} whereas for BFO, the penetration depth is $\sim$10~$\mu$m.~\cite{Talbayev2008} Thus, the laser probe beam in our pump-probe
experiment can see the carrier dynamics at the interface.

\begin{figure}[t] \centering
\includegraphics[width=8cm]{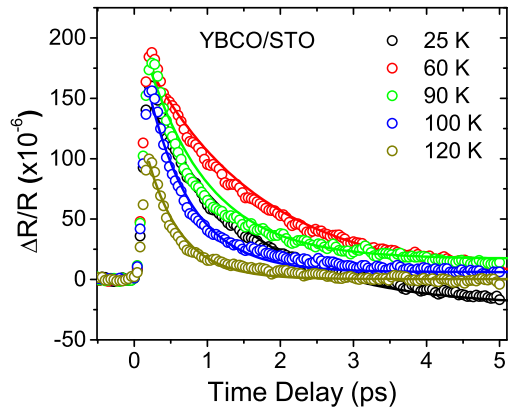}
\caption{Time dependence of photoinduced change in reflectivity ${{\Delta}R}/{R}$ of YBCO thin film. The solid lines show single-exponential fits to the data.}
\label{fig:2a}
\end{figure}

\begin{figure}[t] \centering
\includegraphics[width=8cm]{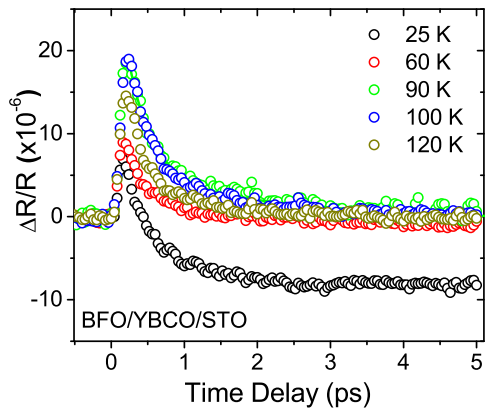}
\caption{Time dependence of photoinduced change in reflectivity ${{\Delta}R}/{R}$ of BFO/YBCO thin film.}
\label{fig:2b}
\end{figure}

\section{Results \& Discussions}
\begin{figure} \centering
\includegraphics[width=8.5cm]{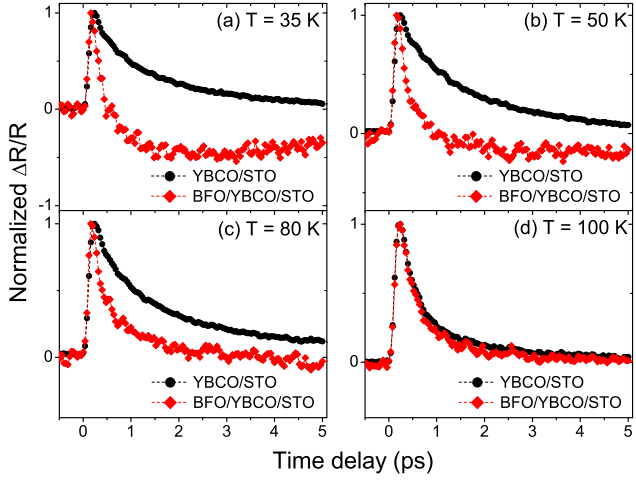}
\caption{Comparison of time dependence of normalized differential reflectivity for the BFO/YBCO and YBCO thin films at (a) 35~K, (b) 50~K, (c) 80~K and (d) 100~K.}
\label{fig:4}
\end{figure}

Comparing the signals, the presence of BFO layer is seen to speed up the relaxation dynamics of YBCO as compared to the pristine YBCO film below  $T_{c}$. Figures~\ref{fig:2a} and ~\ref{fig:2b} show the time dependence of photoinduced change in reflectivity ${{\Delta}R}/{R}$ of YBCO thin film and BFO/YBCO bilayer thin film respectively, at different temperatures below and above $T_{c}$. The YBCO film shows fast initial relaxation dynamics with characteristic relaxation times of $\sim$0.6, 1.0, 1.1 and 0.9~ps at $T$=100, 80, 50 and 35~K respectively. The  few-ps quasiparticle dynamics agrees well with that previously reported.~\cite{Demsar1999} Interestingly, the BFO/YBCO film shows faster relaxation dynamics with characteristic relaxation times of $\sim$0.6, 0.4, 0.3 and 0.3~ps at the corresponding temperatures (for ${{\Delta}R}/{R}$ to decay to half its maximum peak value), indicating a boosted relaxation dynamics in YBCO below $T_c$ in the presence of BFO. To show this speeding up of the relaxation more vividly, we have normalized the ${{\Delta}R}/{R}$ with the peak value and plotted the data of YBCO and BFO/YBCO films comparatively in Fig.~\ref{fig:4}.

The peak value of reflectivity change $|{{\Delta}R}/{R}(T)|$ for YBCO is determined~\cite{Han1990} and plotted in Fig.~\ref{fig:ybco}(a). The temperature
dependent relaxation time $\tau$ of YBCO plotted in Fig.~\ref{fig:ybco}(b) is extracted using single-exponential decay fit to the reflectivity curves. The
amplitude decreases with increasing temperature approaching 90~K, while the relaxation time exhibits an upturn near this temperature. This can be attributed to a phonon bottleneck due to the opening
of a (superconducting) gap in the density of states, as YBCO enters the superconducting state. In order to analyze the temperature dependence of the relaxation amplitude
quantitatively, we use the model proposed by Kabanov \textit{et al}.~\cite{Kabanov1999} The temperature-dependence of the relaxation amplitude $A$ in the
superconducting state for a temperature-dependent gap $\Delta$($T$) is given by:
\begin{equation}
A(T)=\gamma\frac{\epsilon_{I}/(\Delta(T)+k_{B}T/2)}{1+\zeta\sqrt{2k_{B}T/\pi\Delta(T)}\exp[-\Delta(T)/k_{B}T]}.
\label{amp}
\end{equation}
Here $\epsilon_{I}$ is the pump laser intensity per unit cell and $\zeta$ is a constant. The above expression for $A$($T$) describes a reduction in the
photoexcited QP density with increase in temperature, due to the decrease in gap energy and corresponding enhanced phonon emission during the
initial relaxation.

We first fit $|\frac{{\Delta}R}{R}|$($T$) using Eq.~(\ref{amp}) and $|\frac{{\Delta}R}{R}|(T) \propto A$($T$). The result is shown as a solid line in Fig.~\ref{fig:ybco}(a). The fitted value of $T_{c}$, embedded inside the BCS expression for $\Delta (T)$ in Eq.~(\ref{amp}), is
91~K. The fitted value of zero-temperature gap $\Delta$(0)=(2.72$\pm$0.04)$k_{B}T_{c}$ is consistent with the data ($\Delta(0)=3k_{B}T_{c}$) from optical
conductivity studies on YBCO films using terahertz radiation.~\cite{Tsai2003} In the next two sub-sections we use the Rothwarf-Taylor (RT) model~\cite{Rothwarf1967} to fit the data for YBCO and BFO/YBCO.

\subsection{YBCO: Rothwarf-Taylor Model}
\begin{figure} \centering
\includegraphics[width=8.5cm]{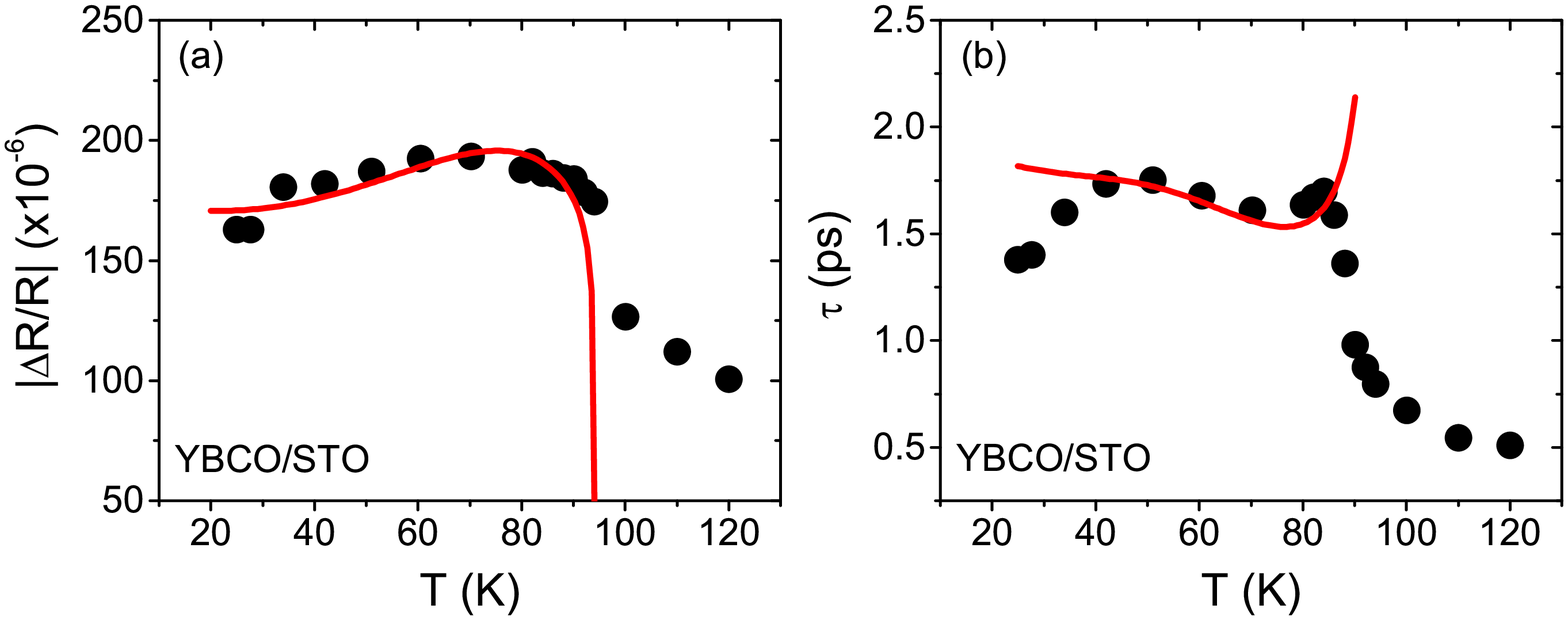}
\caption{(a) Temperature dependence of relaxation amplitude $A(T)$ for YBCO film. Solid line =  fit using Eq.~(\ref{amp}). (b) Temperature dependence
of relaxation time $\tau (T)$ obtained from single-exponential fit (solid circles). Solid line = Fit of $\tau (T)$ using RT model in Eq.~(\ref{tau}).}
\label{fig:ybco}
\end{figure}
\begin{figure}[b]
\centering
\begin{center}
\includegraphics[width=8cm]{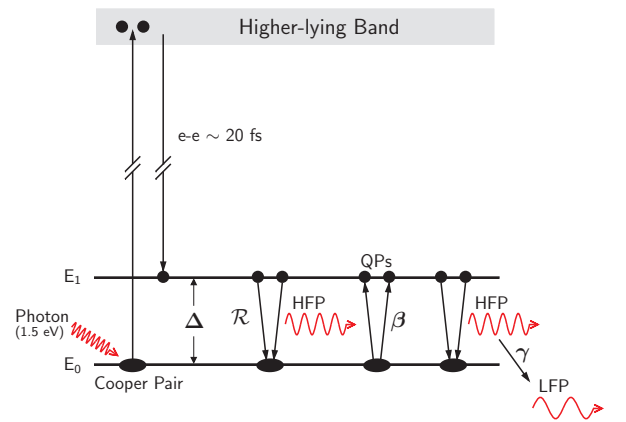}
\caption{Schematic representation of the RT Model. Photoexcited QPs recombine to form Cooper pairs and high-frequency phonons (HFP). The bottleneck for QP relaxation dynamics due to HFP induced reexcitation is governed by the decay rate of HFP into low-frequency phonons (LFP).}
\label{fig:Model}
\end{center}
\end{figure}

The RT model is a phenomenological model that describes the dynamics of photoexcited QPs and high-frequency phonons (HFPs), where the presence of a gap ($\Delta$) in the
electronic density of states gives rise to a bottleneck for carrier recombination (Fig.~\ref{fig:Model}). When two QPs with energies $\geq\Delta$ recombine, a HFP is created with $\omega>2\Delta$. These HFPs trapped within the excited volume can re-break Cooper pairs and act as a bottleneck for QP
relaxation. Hence the superconducting recovery is governed by the decay of the HFP population. The RT model consists of two coupled differential equations
\begin{eqnarray}
\frac{dn_{\text{SC}}}{dt} &=& I_{0} + \beta N - \mathcal{R} n_{\text{SC}}^{2} \nonumber \\
\frac{dN}{dt} &=& \frac{\mathcal{R} n_{\text{SC}}^{2}}{2} - \frac{\beta N}{2}  - \gamma \left(N - N_{\text{Ph}} \right) \ .
\label{RT_Model}
\end{eqnarray}
Here $n_{\text{SC}}$ is the population of the QPs (with energy $E_{1}$ as labeled in Fig.~\ref{fig:Model}) in the YBCO layer and $N$ is the HFP population. The recombination of QPs into Cooper pairs (with probability $\mathcal{R}$) is accompanied by the creation of HFPs with energy $2\Delta$. These HFPs may re-break Cooper pairs with a probability $\beta$ and decay with the rate $\gamma$ (either by anharmonic decay or diffusion out of excitation volume into the substrate).
Based on the RT model, in the superconducting state ($T<T_{c}$), the temperature dependence of the QP relaxation rate $\tau^{-1}$ is determined by the temperature dependence of the relaxation amplitude $A$($T$) and is given by:~\cite{Cao2008,Kabanov2005,Chia2006,Nair2010,ChiaPSS2011}
\begin{multline}
\tau^{-1}(T)=\Gamma\left\{{\delta}A(T)+\eta\sqrt{\Delta(T)T}\exp\left[-\Delta(T)/T\right]\right\}\\
\times\left[\Delta(T)+{\alpha}T\Delta(T)^{4}\right]. \label{tau}
\end{multline}
where $\Gamma$, $\delta$, $\eta$ and $\alpha$ are fitting parameters. 
The solid line in Fig.~\ref{fig:ybco}(b) is obtained from Eq.~(\ref{tau}) and reproduces the upturn of $\tau$ at $T_{c}$ found from single-exponential fits. 

\begin{figure}[t] \centering
\includegraphics[width=8cm]{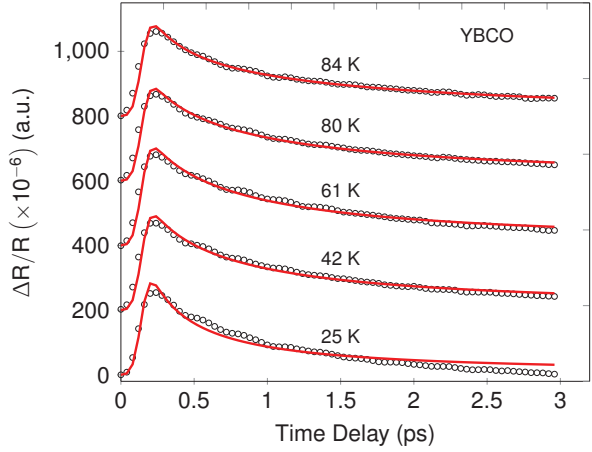}
\caption{Waterfall plots of ${\Delta}R/R$ as function of pump-probe time delay of YBCO thin films at different temperatures. Solid lines are fits to the RT Model (Eq.~(\ref{RT_Model})).}
\label{fig:YBCORTFits}
\end{figure}

\begin{figure}[t]
\centering
\includegraphics[width=8cm]{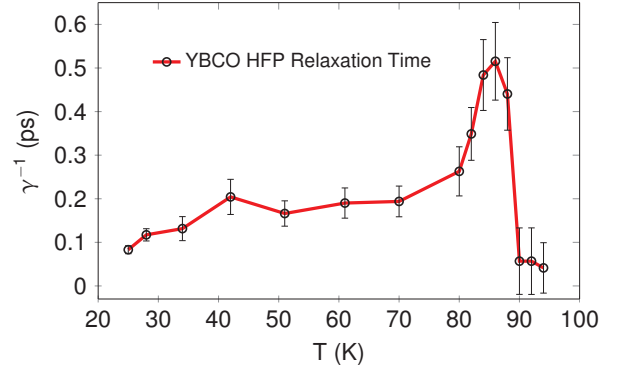}
\caption{Temperature dependence of HFP relaxation time $\gamma^{-1}(T)$ obtained from fits of the numerical solution of the RT model to the experimental $\Delta R/R$ data of pure YBCO.}
\label{fig:TauTrend}
\end{figure}

\begin{figure}[t]
\centering
\includegraphics[width=8cm]{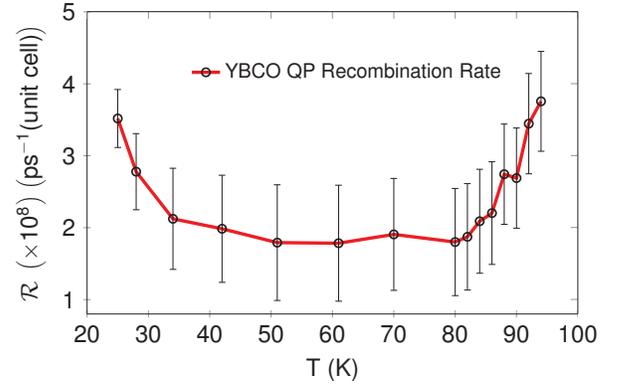}
\caption{Temperature dependence of the QP recombination rate $\mathcal{R}(T)$ obtained from fits of the numerical solution of the RT model to the experimental $\Delta R/R$ data of pure YBCO. Approaching $T_c$ from below, the recombination rate starts to increase, possibly related to the vanishing of the superconducting energy gap $\Delta$. }
\label{fig:RTParameters}
\end{figure}

Next we compare the trend of $\tau$ with the HFP relaxation time $\gamma^{-1}$ in the RT model by directly fitting the RT model to the photoinduced change in reflectivity given by
\begin{equation}
\Delta R/R \propto \Delta n_\text{SC}
\label{dRR}
\end{equation} 
with $\Delta n_\text{SC} = n_\text{SC}^\text{pump} - n_\text{SC}^\text{no pump}$ and the initial condition $n_\text{SC}^\text{no pump}=0$.
From the fits to the RT model (solid lines in Fig.~\ref{fig:YBCORTFits}) we obtain the temperature dependence of $\gamma^{-1}$ as shown in Fig.~\ref{fig:TauTrend}, where we observe an upturn at $T_c$ followed by a sharp drop above $T_c$, similar to Fig.~\ref{fig:ybco}(b). This shows that the relaxation dynamics in the superconducting phase can be explained by the presence of a relaxation bottleneck due to long-lived HFP causing increased reexcitation of QPs.
The downturn of $\tau$ at low temperature is also seen in YBCO single crystals ($T_{c}$=93~K).~\cite{Demsar1999} Apart from the relaxation rate $\gamma$($T$) we obtain the QP recombination rate $\mathcal{R}$($T$) and the pair-breaking rate $\beta$. In the course of fitting we found the temperature dependences and absolute values of $\mathcal{R}$($T$) and $\gamma$($T$) to be independent of pair-breaking rate $\beta$, with no obvious temperature dependence in $\beta$. We find that the best fits are obtained for $\beta=0.1$~ps$^{-1}$.  
Figure~\ref{fig:RTParameters} shows the temperature dependence of the fitted QP recombination rate $\mathcal{R}(T)$, showing $\mathcal{R}$$\sim$$2 \times 10^{8}$~ps$^{-1}$~unit cell in the superconducting state. We compare $\beta$ and $\mathcal{R}$($T$) with those obtained for the conventional superconductor MgB$_{2}$,~\cite{Demsar2003} where $\mathcal{R}$$\sim$$1 \times 10^{2}$~ps$^{-1}$~unit cell and $\beta$$\sim$$7 \times 10^{-2}$~ps$^{-1}$. We attribute the differences to the very different QP lifetimes between YBCO ($\sim$1--2~ps) and MgB$_{2}$ ($\sim$300~ps), caused by the larger superconducting gap value in YBCO, that results in the dynamics in YBCO being governed by the optical phonons, instead of acoustic phonons in MgB$_{2}$.~\cite{Demsar2003} Moreover we observe, that near $T$$\sim$$80$~K, and hence still within the superconducting regime, the recombination rate starts to increase. This behavior can be associated with the rapid closing of the energy gap $\Delta$ resulting in a reduced population time for the gap edge state. 

In the next sub-section we analyze the BFO/YBCO data. The obvious speeding up of the YBCO dynamics (or decrease in the relaxation time) in the BFO/YBCO bilayer, compared to the pure YBCO film (see Fig.~\ref{fig:4}), may be attributed to the presence of the BFO overlayer. Theoretically it has been shown that the BFO overlayer near the interface can become a ferromagnetic metallic state.~\cite{Zhu:2014aa} The BFO overlayer could either act as a trapping layer for QPs or, alternatively, the weak ferromagnetism of the BFO at the BFO/YBCO interface could interact with the superconductivity in the YBCO, resulting in a speeding up of the decay of HFPs to low-frequency phonons, and hence the speeding up of the QP relaxation in YBCO. The presence of the interfacial weak ferromagnetic state necessitates the use of an extended Rothwarf-Taylor model.

\subsection{BFO/YBCO: Extended Rothwarf-Taylor Model}
\begin{figure}[t]
\centering
\includegraphics[width=8cm]{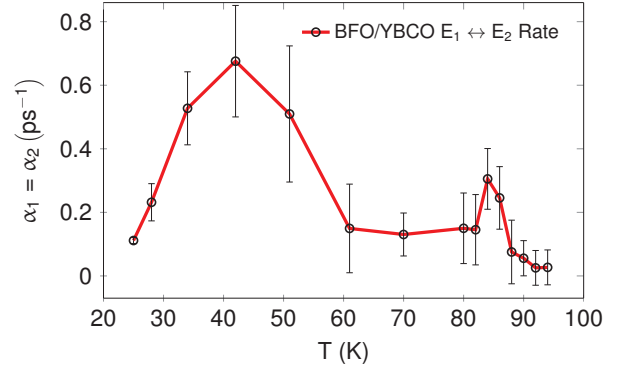}
\caption[]{The excitation and relaxation rates ($\alpha_1, \alpha_2$) between the gap edge state ($E_1$) and the ferromagnetic interfacial layer ($E_2$) show a peak close to $T_c$$\sim$84~K. This indicates that more QPs are trapped in the interfacial state $E_2$ and this in turn causes a reduction of QP recombinations. The peak in $\alpha$ coincides with the closing of the energy gap (see Fig.~\ref{fig:RTParameters}). }
\label{fig:E1E2Rate}
\end{figure}

To describe the BFO/YBCO data we extend the original RT model by an additional energy level ($E_2$) above the gap edge ($E_1$). We propose this level $E_{2}$ originates from the weak ferromagnetic interfacial layer. Photoexcited QPs transiently occupy $E_{2}$, then relax to the gap-edge state $E_{1}$, before they undergo recombination into Cooper pairs (Fig.~\ref{fig:Extended_Model}). The differential equations for this extended model are of the form:
\begin{figure}[t]
\centering
\includegraphics[width=8cm]{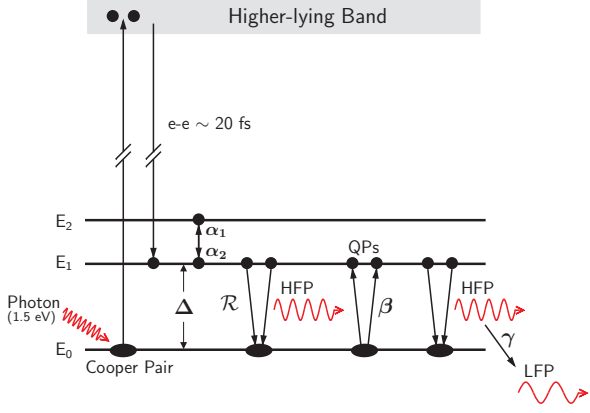}
\caption{Schematic of extended RT Model, where an additional energy level $E_2$ is present due to the weak ferromagnetic interfacial layer. LFP = low-frequency phonon.}
\label{fig:Extended_Model}
\end{figure}

\begin{eqnarray}
\frac{dn_{\text{SC}}}{dt} &=& I_{0} + \beta N - \mathcal{R} n_{\text{SC}}^{2} - \alpha_{1} n_{\text{SC}} + \alpha_{2} n_{\text{Int}} \nonumber \\
\frac{dn_{\text{Int}}}{dt} &=&  \alpha_1 n_{\text{SC}} - \alpha_{2} n_{\text{Int}} \nonumber \\
\frac{dN}{dt} &=& \frac{\mathcal{R} n_{\text{SC}}^{2}}{2} - \frac{\beta N}{2}  - \gamma \left(N - N_{\text{Ph}} \right)\ .
\label{ExtRT_Model}
\end{eqnarray}
Compared to the original RT model, $n_{\text{Int}}$ is the population of QPs (with energy $E_{2}$) at the BFO/YBCO interface. The parameters $\alpha_1$ and $\alpha_2$ represent the coupling between the QPs at the YBCO and BFO/YBCO interface. This coupling may result from the $d$ electrons of the YBCO CuO$_{2}$ planes/chains hybridizing with the $d$ electrons in the BFO, at the interface. This local hybridization can then be mapped to the hybridization of Bloch states in the YBCO and interfacial layer. The photoinduced change in reflectivity is given by

\begin{equation}
\Delta R/R = \sigma_1 \Delta n_{\text{SC}} + \sigma_2 \Delta n_{\text{Int}} \ ,
\label{dRR_Ext}
\end{equation}
again with $\Delta n_{i} = n_{i}^\text{pump} - n_{i}^\text{no pump}$ where $\sigma_1 > 0$ corresponds to the pure YBCO case in Sec. III.A, and $\sigma_2 < 0$ accounts for the effect of the interfacial QPs on $\Delta R/R$. Since the weak ferromagnetic state is expected to exist only within a few unit cells of the BFO/YBCO interface,~\cite{Zhu:2014aa} the bulk of the YBCO layer (100~nm thick) should be be identical to that of the pristine YBCO film. Hence we use the same values of $\beta$, $\mathcal{R}$($T$) and $\gamma$($T$) in our fittings of the BFO/YBCO data, and only vary $\sigma_{1}$, $\sigma_{2}$, $\alpha_{1}$ and $\alpha_{2}$. We also make the physically reasonable assumption that $\alpha_{1} = \alpha_{2}$.  Figure.~\ref{fig:E1E2Rate}, which shows the fitted $\alpha$($T$), reveals a peak near $T_c$ indicating increased population of the additional energy level $E_2$. This implies a reduction of the total number of QP recombinations. The peak at $T$$\sim$84~K coincides with the temperature where the recombination rate $\mathcal{R}$($T$) starts to increase (Fig.~\ref{fig:RTParameters}). 
 
\begin{figure}[t]
\centering
\includegraphics[width=8cm]{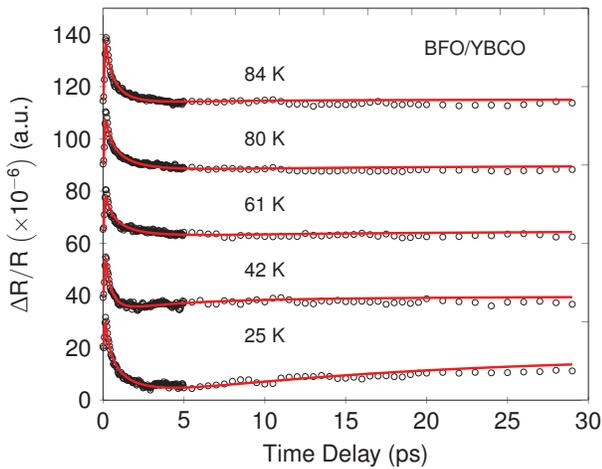}
\caption[]{Waterfall plots of ${\Delta}R/R$ as function of pump-probe time delay of BFO/YBCO thin films at different temperatures. 
Solid lines are fits to the extended RT Model (Eq.~(\ref{ExtRT_Model})).}
\label{fig:ExtendedRT}
\end{figure}

 
Figure~\ref{fig:ExtendedRT} shows that the BFO/YBCO data can be fitted successfully using the extended RT Model [Eq.~(\ref{dRR_Ext})]. The model reproduces the low-temperature fast relaxation and crossing of ${\Delta}R/R$ from positive to negative at short time delays, as well as the recovery at long time delays. In the recent pump-probe work of LSMO/YBCO bilayer, a similar fast initial relaxation, as well as a zero crossing of ${\Delta}R/R$ were also observed, compared to pure YBCO.~\cite{Parlato2013} The authors attributed these features to the appearance of an ultrathin YBCO layer at the LSMO-YBCO interface, which is "depressed" due to its proximity with the ferromagnetic LSMO layer, and acts as an energy trap (suppressed 2$\Delta$ region) that shortens the relaxation process. This lends credence to our assertion that $E_{2}$ arises from the weak ferromagnetic interfacial layer in our BFO/YBCO bilayer.

\section{Conclusion}
We have performed optical pump-probe reflectivity measurements on YBCO and BFO thin films, and on BFO/YBCO bilayer thin film heterostructure. The results show that superconductivity in YBCO is retained in the heterostructure. Compared with pure YBCO, the presence of BFO layer on top of the superconducting YBCO speeds up the fast relaxation in YBCO, but only below $T_{c}$. We attribute this fast relaxation to the presence of weak ferromagnetism at the BFO/YBCO interface. From fitting numerically obtained solutions of the RT model to the experimental YBCO data, we observe a peak in the HFP relaxation time $\gamma^{-1}$. This is in agreement with a peak in the QP relaxation time $\tau$ in single exponential fits. To fit the $\Delta R/R$ data for BFO/YBCO we use the parameter trends obtained from fitting the YBCO data. The difference between the YBCO and BFO/YBCO $\Delta R/R$ data can be described by adding a ferromagnetic interfacial layer with $E_2>E_1$ into the original RT model. For the transition rates between the energy levels $E_1$ and $E_2$ we find a peak near $T_c$. This corresponds to an increased occupation of the interfacial layer and hence a reduced number of QP recombinations near $T_c$.
These results may further stimulate the use of pump-probe techniques for studies on other heterostructures to explore the interactions across the interfaces and to construct advanced devices with tunable functionalities.

\section{Acknowledgements}
E.E.M.C. thanks Jure Demsar for useful discussions. The work at NTU was supported by Singapore MOE AcRF Tier 1 (RG 13/12 \& RG123/14) and National Research Foundation Competitive Research Programme (NRF-CRP4-2008-04).  The work at Los Alamos (J.-X.Z.) was carried out under the auspices of the NNSA of the US DOE at LANL under Contract No. DE-AC52-06NA25396, and was supported by the LANL LDRD Program. The work was supported in part by the Center for Integrated Nanotechnologies, a U.S. DOE BES user facility.

\bigskip

\bibliography{BYS}
\end{document}